\documentclass[12pt]{article}
\usepackage{amssymb}

\newcommand{\pixi}{\ensuremath{\pi^R_{\tau}({\vec{\xi}})  }  }
\newcommand{\phixi}{\ensuremath{\phi^R_{\tau}({\vec{\xi}})  }  }

\newcommand{\pit}{\ensuremath{\pi_{t}({\bf{x}})  }  }
\newcommand{\phit}{\ensuremath{\phi_{t}({\bf{x}})  }  }
\newcommand{\adx}{\ensuremath{{a_t}^{\dagger} (\bf{{x}})}}
\newcommand{\ax}{\ensuremath{a_t ({\bf{x}})}}

\newcommand{\adk}{\ensuremath{{a_t}^{\dagger} ({\bf{k}})}}
\newcommand{\ak}{\ensuremath{a_t ({\bf{k}})}}

\newcommand{\bdxi}{\ensuremath{{b^R_\tau}^{\dagger} (\vec{{\xi}})}}
\newcommand{\bxi}{\ensuremath{b^R_\tau ({\vec{\xi}})}}

\newcommand{\bdk}{\ensuremath{{b^R_\tau}^{\dagger} ({\vec{\kappa}})}}
\newcommand{\bk}{\ensuremath{b^R_\tau ({\vec{\kappa}})}}

\newcommand{\Htau}{\ensuremath{{\cal{H}}_{\tau}} \,}

\newcommand{\V}{\ensuremath{{\cal{V}} \, }}
\newcommand{\B}{\ensuremath{{\cal{B}} \, }}
\newcommand{\D}{\ensuremath{{\cal{D}} \, }}
\newcommand{\Z}{\ensuremath{{\cal{Z}} \, }}
\newcommand{\Ct}{\ensuremath{{\cal{C}}_t \, }}
\newcommand{\M}{\ensuremath{{\cal{M}} \, }}
\newcommand{\cH}{\ensuremath{{\cal{H}} \, }}
\newcommand{\F}{\ensuremath{{\cal{F}} \, }}
\newcommand{\R}{\ensuremath{{\cal{R}} \, }}
\newcommand{\Ctau}{\ensuremath{{\cal{C}}_\tau^\R \, }}

\newcommand{\cL}{\ensuremath{{\cal{L}} \, }}

\begin{document}

\renewcommand{\thefootnote}{\fnsymbol{footnote}}

\begin{titlepage}

\title{A Consistent Histories approach to the Unruh Effect}
\author{Duncan Noltingk\footnotemark[1] \\
        \\
        \emph{Blackett Laboratory}\\
        \emph{Imperial College}\\
        \emph{Prince Consort Road}\\
        \emph{London SW7 2BZ}}

\maketitle

\footnotetext[1]{d.noltingk@ic.ac.uk}

\begin{abstract}
Using the HPO approach to consistent histories we re-derive
Unruh's result that an observer constantly accelerating through
the \newline Minkowski vacuum appears to be immersed in a thermal bath. We
show that propositions about any symmetry of a system always form
a consistent set and that the probabilities associated with such
propositions are decided by their value in the initial state. We
use this fact to postulate a condition on the decoherence
functional in the HPO set-up. Finally we show that the Unruh
effect arises from the fact that the \emph{initial} density matrix
corresponding to the inertial vacuum can be written as a thermal
density matrix in the Fock basis associated with the accelerating
observer.
\end{abstract}

\end{titlepage}

\section{Introduction}

\subsection{Consistent Histories}

 The consistent histories approach to quantum theory originated in the
pioneering work of Griffiths \cite{Gri} and Omnes \cite{Omn}.
Initially the formalism was developed in an attempt to escape the familiar
difficulties of the Copenhagen interpretation. More recently, Gell-Mann and
Hartle \cite{GH} suggested that generalised history theories may
be useful in tackling the problems of quantum cosmology and
quantum gravity, in particular the problem of time.

 The basic ingredient of `conventional' consistent histories is a
time-ordered sequence of propositions about the system represented
by a class operator:
\begin{equation}
  C_\alpha := \alpha_{t_1}(t_1) \alpha_{t_2}(t_2) \cdots \alpha_{t_n}(t_n)
\end{equation}
 where $\alpha_{t_i}(t_i)$ is a Heisenberg picture projection operator
representing a proposition made about the system at time $t_i$. To
make physical predictions we must use the decoherence functional
to identify (strongly) consistent sets of
 histories, \emph{i.e.} sets $\{\alpha_i\}$ such that,
\begin{eqnarray}
        \label{decfun}
  d(\alpha_i,\alpha_j) &:=& Tr_{\cal{H}}[C_{\alpha_i}^\dagger \rho
                                       \, C_{\alpha_j}]    \\
                       &=& 0 \, \, \mbox{if} \,\, i \, \neq \, j 
\end{eqnarray}
 Within such consistent sets, the probability of a particular history
$\alpha_i$ `occuring' is $d(\alpha_i,\alpha_i)$. The consistency
condition guarantees that the Kolmolgorov sum rules are satisfied.
 
If generalised history theories are to be useful in formulating quantum 
gravity, then it is important to understand how more conventional theories such
as non-relativistic quantum mechanics and quantum field theory (QFT) can be 
formulated in history language. While non-relativistic quantum mechanics has
been extensively studied within the formalism, there are very few results 
concerning QFT. This is the motivation for this paper in which we re-derive a 
well-known result in the theory of QFT on curved spaces, from a histories
perspective. The Unruh effect \cite{Unruh1} is an analogue of Hawking 
radiation, but the gravitational field that induces the radiation is `apparent'
rather than `real', \emph{i.e.} it is measured by an observer accelerating 
through empty space rather than by an observer in
the gravitational field of a black hole.

\subsection{The HPO Approach}

 Isham \cite{Ish1} proposed an algebraic scheme for generalised history
theories of the type suggested by Gell-Mann and Hartle. The
algebraic axioms are set up in analogy with the logical approach
to single time quantum theory which is concerned with the pair
$(\cal{L},\cal{S})$ where $\cal{L}$
is the lattice of projection operators on a Hilbert space and
$\cal{S}$ is the set of density matrices. Isham proposed that a
generalised history theory should be composed of the pair
$(\cal{U}\cal{P},\cal{D})$ where $\cal{U}\cal{P}$ is an
\emph{orthoalgebra} of propositions about possible histories and
$\cal{D}$ is the space of decoherence functionals.

To fit conventional consistent histories into these axioms,
we would like to interpret the class operators as logical propositions;
however, the product of non-commuting projection operators is not 
a projection operator. This means it is difficult to
define conjuctions, disjunctions and negations consistently.
 However, the \emph{tensor product} of two projectors \emph{is} a projector
on the tensor product space.  This is the central idea of the
history projection operator (HPO) approach to consistent
histories. The tensor product of Schr\"odinger picture projection
operators, $\alpha_{t_1} \otimes \alpha_{t_2} \otimes \cdots
\otimes \alpha_{t_n}$, which is a projector on the n-time history
space, $\V^n := {\cal{H}}_{t_1} \otimes {\cal{H}}_{t_2} \otimes
\cdots \otimes {\cal{H}}_{t_n}$, represents the proposition
``$\alpha_{t_1}$ is true at time $t_1$ \emph{and then}
$\alpha_{t_2}$ is true at time $t_2$ $\dots$ \emph{and then}
$\alpha_{t_n}$ is true at
 time $t_n$.'' Now we can define the logical operations as we would for
projection operators in any Hilbert space. So in this case, the
orthoalgebra  $\cal{U}\cal{P}$ is in fact the lattice
${\cal{P}}({\cal{V}}^n)$ of projection operators on the history
space.

The decoherence functional (\ref{decfun}) can be written as
\begin{equation}
  d(\alpha_i,\alpha_j) = Tr_{\V^n \otimes \V^n}(\alpha_i \otimes \alpha_j X)
\end{equation}
 for some $X \in \B(\V^n \otimes \V^n)$ where $\B({\cal{H}})$ is defined as 
the set of bounded operators on $\cal{H}$. Conversely 
Gleason's theorem can be used to show that any decoherence functional
that satisfies certain natural conditions can be written in this form
\cite{Ish2}. Therefore \D, the space of decoherence functionals,
is the set of all functionals of this form. This result also holds
in the continuous time case \cite{Ish3}.

\newpage

\section{The Simple Harmonic Oscillator}

\subsection{Continuous Times}

 In extending HPO theory to the case of continuous time, which we anticipate to
be important for QFT, we encounter the
continuous tensor product of the single-time Hilbert space:
$\V^{cts} := \otimes_{t \in \mathbb{R}} {\cal{H}}_t$. To deal with
this object it is useful to confine ourselves for the moment to
the simple harmonic oscillator (SHO), where ${\cal{H}}_t =
L^2(\mathbb{R})$ and to consider the \emph{history group} \cite{Ish3}. We can 
view $\V^n$ arising as the representation space for the n-fold direct
product of the Weyl group of single time quantum theory on the
line:
\begin{eqnarray}
  \left[ x_{t_i} , x_{t_j} \right] &=& 0 \\
  \left[ p_{t_i} , p_{t_j} \right] &=& 0 \\
  \left[ x_{t_i} , p_{t_j} \right] &=& i\hbar\delta_{ij}
\end{eqnarray}
 The advantage of this perspective is that it can be readily generalised to the
case of continuous time. For this we consider the algebra:
\begin{eqnarray}
        \label{SHOalg}
  \left[ x_f , x_g \right] &=& 0 \\
  \left[ p_f , p_g \right] &=& 0 \\
  \left[ x_f , p_g \right] &=& i\hbar(f,g)
\end{eqnarray}
where $f,g \in L^2(\mathbb{R})$ ; $x_f := \int dt f(t)x_t$ and
$(f,g) := \int dt \, f(t) g(t)$. This algebra is clearly
isomorphic to the algebra of a one-dimensional QFT and
 suggests that field theory techniques will be useful in
studying the theory. It is well-known that this algebra has a
representation on the Fock space over $L^2(\mathbb{R})$, denoted
$\F[L^2(\mathbb{R})]$. Indeed it can be shown that \cite{Ish3},
\begin{equation}
 \label{iso}
 \V^{cts} := \otimes_{t \in \mathbb{R}} {\cal{H}}_t \approx \F[L^2(\mathbb{R})]
\end{equation}
and again  $\cal{U}\cal{P}$ is a lattice, now it is the set
of projection operators on the continuous history space,
${\cal{P}}({\cal{V}}^{cts})$.
 The condition that the time-averaged Hamiltonian is self-adjoint is sufficient
to select a unique representation of the history algebra \cite{Ish3}
. This representation is defined by the Fock basis associated with
the creation operator,
\begin{equation}
 a_f^\dagger := \sqrt{\frac{m\omega}{2\hbar}}x_f - 
				i\sqrt{\frac{1}{2m\omega\hbar}}p_f
\end{equation}

\subsection{Time averaged propositions}

 The physical interpretation of a continuous time HPO theory is based on the 
assumption that
projectors onto the spectrum of self-adjoint operators on \V
represent propositions about the time-averages of physical
quantities. So projections onto the eigenvectors of the $x_f$
operators introduced in (\ref{SHOalg}) represent propositions
about the average position of the particle over time. As $[x_f ,
x_g] = 0$, these operators have common eigenvectors for any
smearing function. We denote these eigenvectors $| x(\cdot)
\rangle$ and they can be interpreted as fine-grained histories or
trajectories of the particle. In the single-time theory, $x_t |x
\rangle = x |x \rangle$ so, formally, $x_t| x(\cdot) \rangle =
x(t)| x(\cdot) \rangle$, which suggests;
\begin{equation}
  x_f | x(\cdot) \rangle := \int dt \, f(t) x_t | x(\cdot) \rangle
                = (f,x)| x(\cdot) \rangle
\end{equation}
 If this is to make sense then $x(\cdot)$ must be a member of 
$L^2(\mathbb{R})$. However, it is likely that the eigenvectors $x(\cdot)$ will 
be distributions rather than functions. The natural procedure now would be to 
interpret the symbol $(f,x)$ to be the real
number obtained from the pairing of the distribution $x$ with the function $f$.
This implies that the allowed functions $f$ should really be members of 
Schwartz space rather than $L^2(\mathbb{R})$. We will not confront this issue
here, and just consider functions which are members of some unspecified space,
$\tau$.

 For each $f \in \tau$ we have an equivalence relation, $\sim_f$, on
trajectories if we define $x(\cdot) \sim_f y(\cdot)$ if $(f,x) =
(f,y)$. We denote these equivalence classes by $[(f,x)]$.
 Now we consider projections onto the spectrum of $x_f$. We denote the
operator which projects onto the eigenvector of $x_f$ with
eigenvalue $(f,x)$ as $P_{(f,x)}$; it projects onto the
equivalence class of trajectories $[(f,x)]$, \emph{i.e.} onto a
coarse-grained history.
 Similar remarks obviously apply to operators $P_{(f,p)}$ which project onto
coarse grained momentum trajectories $[(f,p)]$.

 Another operator of physical significance is the smeared Hamiltonian:
\begin{eqnarray}
  H_f &:=& \int dt \, f(t) (\frac{1}{2m}p_tp_t +
		\frac{m\omega^2}{2} x_tx_t)  \\
      &=&  \hbar\omega \! \int dt \, f(t) \, (a_t^\dagger a_t
			+ \frac{1}{2})
\end{eqnarray}
Projections onto its spectrum represent propositions about the time-averaged
energy of the system.

 For our purposes, the average number operator $N$ will be of prime importance.
We can formally define it as follows:
\begin{equation}
  N := \int dt \, a^\dagger_t a_t
\end{equation}
 The eigenvectors of this operator are vectors of the form
\begin{equation}
 |n_f \rangle :=  (n!)^{-1/2} \int dt f(t_1 \dots t_n) a_{t_1}^\dagger \dots
                                        a_{t_n}^\dagger |0\rangle
\end{equation}
 These are also eigenvectors of the Hamiltonian.
 The average number operator has a highly degenerate spectrum as vectors of the
above form have eigenvalue $n \in \mathbb{N}$ for all smearing functions $f$, 
as can be easily checked.
 We will denote the projection operator onto $|n_f \rangle$ as $P_{n_f}$; it
represents a proposition about the average number of quanta present in a
particular time interval.

\subsection{Propositions within a finite time interval}

 We can write the average number operator defined above in the form,
$N = N_{f=1}$ where $N_f := \int dt f(t) a_t^\dagger a_t$.
This shows that there is a problem with the
definition because the constant function $f=1$ is not a member of
$L^2(\mathbb{R})$. However it is a member of $L^2[a,b]$ where $[a,b]$ is a
finite interval of the real line. This suggests that we should really be
dealing with propositions in a finite interval of time.

 Consider again the proposition $P_{n_f}$. Intuitively the
support of $f$ affects the time period in which the proposition is made.
In other words if $supp(f) \subset [a,b]$ then the proposition $P_{n_f}$ refers
to the average number of particles during the time period $[a,b]$.
 We can formulate this rigorously by splitting up ${\cal{V}}^{cts}$ as follows:
\begin{eqnarray}
  {\cal{V}}^{cts} &:=& \otimes_{t \in \mathbb{R}} {\cal{H}}_t \\
        &=& \V^{[-\infty,a]} \otimes \V^{[a,b]} \otimes \V^{[b,\infty]}
\end{eqnarray}
where $\V^{[a,b]} := \otimes_{t \in [a,b]} {\cal{H}}_t$. Now we can use the
isomorphisms:
\begin{equation}
    \label{iso2}
\otimes_{t \in [a,b]}e^{L^2_t[a,b]} \approx e^{{}^\oplus \!\!\! 
		\int_a^b L^2_t[a,b]} \approx \F[L^2[a,b]]
\end{equation}
\cite{Gui}. Here, ${}^\oplus \!\!\! \int_a^b L^2_t[a,b]$ is the direct
integral Hilbert space over the interval $[a,b]$. An element of
this Hilbert space, $F$, can be considered as a one-parameter
family of elements of $L^2[a,b]$ which we denote by $f_t$, where
$t \in [a,b]$. The inner product is defined as
\begin{equation}
 (F,G)_{{}^\oplus \!\!\! \int_a^b L^2_t[a,b]} :=
			\int_a^b dt \, (f_t,g_t)_{L_t^2[a,b]}
\end{equation}
From the right hand side of (\ref{iso2}) we can see that
$\V^{[a,b]}$ naturally carries a representation of the Lie algebra
\begin{eqnarray}
        \left[ x_f , x_g \right] &=& 0          \\
        \left[ p_f , p_g \right] &=& 0  \\
        \left[ x_f , p_g \right] &=& i(f,g)
\end{eqnarray}
 where $f,g \in L^2[a,b]$. The natural interpretation of these operators is
that they are associated with time averaged propositions about position and
momentum in the finite time interval $[a,b]$. We can form complex
combinations of these operators in the usual way to define creation and
annihilation operators. Projections onto the eigenvectors of the average number
 operator associated
with these correspond to propositions about the average number of particles
in the time interval $[a,b]$.

 We can
now see that propositions on ${\cal{V}}^{cts}$ smeared by functions in a
finite time interval are isomorphic with propositions on ${\cal{V}}^{[a,b]}$ by
\begin{equation}
 P_{n_f} \approx \mathbb{I}_{\V^{[-\infty,a]}} \otimes P^{[a,b]}_{n_f} \otimes
                                        \mathbb{I}_{\V^{[b,\infty]}}
\end{equation}
where $f \in L^2[a,b]$ and $P^{[a,b]}_{n_f} \in \V^{[a,b]}$. So from now on
when
we use the average number operator $N$ it should be understood that in fact we
are averaging over a finite time interval \emph{i.e.} we are smearing with
functions $f \in L^2[a,b]$.

 This is consistent with the definition of finite time interval projectors for
 coherent states given by Isham \emph{et al} \cite{Ish3}.

\subsection{The Decoherence Functional}

Isham \emph{et al} \cite{Ish3} and Anastopolous \cite{anas} have
defined decoherence functionals for continuous time projectors in
the HPO scheme by considering projections onto coherent states.
However, we are interested in propositions concerning the average
number of quanta. These cannot be simply related to coherent
states, so we will take a different approach and require our
decoherence functional to respect the dynamical time translation
symmetry of quantum theory. As the projectors onto eigenstates of
$N$ commute with the Hamiltonian, we would expect the probability
of any such proposition to be decided by its probability in the
initial state. We will see that this is indeed the case and that
these propositions also form a `canonical' consistent set.
 We shall then require these conditions to hold in the HPO formalism to obtain
a condition on the decoherence functional. Analogous remarks apply
to any symmetry of the system, \emph{i.e.} propositions regarding
the spectral projectors of any operator which commutes with the
Hamiltonian will form a consistent set and their probabilities
will be decided by their value in the initial state.

 Let us first examine the discrete time case for the SHO with
single-time number operator defined by $N^{st} := a^\dagger a$.
Here we have time translation symmetry $[H,N^{st}] = 0$, which
corresponds to the conservation of the number of quanta.
 We begin by considering a 2-time history in the conventional set-up. It has
eigenvectors $|n\rangle := (n!)^{-1/2}(a^\dagger)^n|0\rangle$ and we denote
Schr\"odinger picture projectors onto these vectors by $P_n$. The
class operator takes a particularly simple form,
 $C_{n_1n_2} := P_{n_1}(t_1)P_{n_2}(t_2) = P_{n_1} P_{n_2} =
\delta_{n_1 n_2}P_{n_1}$. The decoherence functional is then,
\begin{eqnarray}
 d_{SHO}(m_1m_2,n_1n_2) &:=& Tr_\cH [C_{m_1m_2} \rho C^\dagger_{n_1n_2}]  \\
                  &=& \delta_{m_1 m_2} \delta_{n_1 n_2} Tr_\cH[P_{m_1} \rho
                                                                P_{n_1}] \\
                  &=& \delta_{m_1 m_2} \delta_{n_1 n_2} \delta_{m_1 n_1}
                                                        \rho_{m_1 n_1}
\end{eqnarray}
 We can see that the fact that the projectors commute with the Hamiltonian
means that they must all project onto the same state for the
answer to be non-zero. This shows that propositions about the
average number of particles,or more generally propositions about
any symmetry of a system, always make up a consistent set. It is
also clear that the probabilities assigned to these propositions
depend on the initial state alone.

 Now we examine this in the HPO scheme. The history space is
$\V^2 = {\cal{H}}_{t_1} \otimes {\cal{H}}_{t_2}$. We can write the
above decoherence functional as a trace over $\cH^{\otimes5} :=
\cH_{t_0} \otimes \cH_{t_1}\otimes \cH_{t_2} \otimes \cH_{t_1}
\otimes \cH_{t_2}$ using the trick in \cite{Ish3}:
\begin{equation}
  d_{SHO}(m_1m_2,n_1n_2) = Tr_{\cH^{\otimes 5}}[\rho \otimes P_{m_1} \otimes
                        P_{m_2} \otimes P_{n_1} \otimes P_{n_2} S_5]
\end{equation}
 Tracing over the initial Hilbert space we obtain,
\begin{equation}
  d_{SHO}(m_1m_2,n_1n_2) = Tr_{\V^2 \otimes \V^2}[P_{m_1} \otimes P_{m_2}
                        \otimes P_{n_1} \otimes P_{n_2} Z]
\end{equation}
 where $Z \in {\cal{B}}(\V^2 \otimes \V^2)$ and is defined in terms of its
matrix elements in the energy basis as,
\begin{equation}
  \langle i_1 \dots i_4 | Z | j_1 \dots j_4 \rangle = \delta_{i_1j_2}
                                \delta_{i_2j_3}\delta_{i_3j_4}\rho_{i_4j_1}
\end{equation}
 Now it is the operator $Z$ that contains the initial
conditions and forces all the projectors to project onto the same state.
In fact,
 by using these energy eigenstates we have removed the dynamics from the
decoherence functional and are left only with the initial
conditions and temporal structure encoded in the operator $Z$.
Note that this does not uniquely define $Z$ as any $Z'$ defined by
$Z'=  U^\dagger Z U$ where $U$ is of the form $e^{if(H)} \otimes
e^{ig(H)}$ has the same matrix elements if $H$ is the
time-averaged Hamiltonian; $H := \int \, dt H_t$ .

 Consider a continuous time energy proposition in standard history
 theory, represented by the class operator
$C_{\{n_t\}} := \Pi_t P_{n_t}(t)$. Heuristically, this is going to
be zero unless all of the $n_t$ are equal.
 If they are all equal, to $n$ say, then the infinite product will equal $P_n$.
 In this case the decoherence functional will give the same result as before:
\begin{eqnarray}
 d_{SHO}(\{m_s\} \,,\, \{n_t\}) &=& \delta_{mn}\rho_{mn} \,\,
                if \, \, m_s = m \, \forall s \, , \, n_t=n \, \forall t  \\
                     &=& 0 \, \, \mbox{otherwise}
\end{eqnarray}
 We can now understand the degeneracy in the spectrum of the average number
 operator in the HPO approach. It
corresponds to the fact that the number of quanta is conserved and
must be an integer. Therefore the time-averaged number of quanta must be an
integer over any time period.

From the above discussion we require that the continuous time HPO
decoherence functional satisfies
\begin{equation}
 d_{SHO}^{cts}(n_f , m_g) = \delta_{mn}\rho_{mn}
\end{equation}
for all functions $f,g$. This guarantees that:

\medskip

1. The functional $d_{SHO}^{cts}$ assigns the correct
probabilities to average number propositions. $P_{n_f}$
corresponds to the proposition ``There are an average of $n$ quanta
over the time interval $t \in supp(f)$''. However, we know that the
number of quanta is constant in time so the smearing function is
irrelevant and that the probability of finding $n$ particles at
any time is simply $\rho_{nn}$.

2. Number propositions still form a consistent set.

\medskip

There is a class of operators  $Z^{cts} \in
{\cal{B}}({\cal{V}}^{cts} \otimes {\cal{V}}^{cts})$ such that the decoherence
functional can be written in the form:
\begin{equation}
  d_{SHO}^{cts}(n_f , m_g) := Tr_{{\cal{V}}^{cts} \otimes {\cal{V}}^{cts}}
                                        [P_{n_f} \otimes P_{m_g} Z^{cts} ]
\end{equation}
 such $Z^{cts}$ must satisfy,
\begin{equation}
    \label{Zcts}
  \langle m_f n_g | Z^{cts} | m'_{f'} n'_{g'} \rangle := \delta_{mn}
\delta_{nm'}\delta_{m'n'} \rho_{mn}
\end{equation}
for all functions $f, f', g, g'$
 as can be easily shown by taking the trace over energy eigenstates:
\begin{equation}
   Tr_{{\cal{V}}^{cts} \otimes {\cal{V}}^{cts}}[ X ] =
        \int \D \mu[m_f] \D \mu[n_g] \langle m_f n_g |X| m_fn_g \rangle
\end{equation}
 The measure $\D \mu[m_f]$ can be assumed to exist because there is a
well-defined measure on $\V^{cts}$ defined in terms of coherent states
\cite{Ish3}.
 The condition (\ref{Zcts}) only defines $Z^{cts}$ up to a unitary
transformation.

\section{Quantum Field Theory}

\subsection{The HPO approach to QFT}

We use throughout the signature $(+,---)$. To construct an HPO
version of canonical QFT on Minkowski space-time, \M, we must
first foliate \M with a one parameter family of space-like
surfaces using some timelike vector $n^\mu$, normalised by
$\eta_{\mu\nu}n^\mu n^\nu=1$. Note that this corresponds to a
choice of time direction as seen by some inertial observer. This
choice obviously breaks Lorentz covariance and an important
unsolved problem in the HPO programme is to show the equivalence
of theories based on all such slicings. See \cite{Ish4} for a
relevant discussion. In this paper however, we will not consider
this problem and just consider slices orthogonal to the vector $n
:= \partial_{x^0}$ where $x^\mu$ is the coordinate system on \M in
which our inertial observer is at rest. Now we consider a
canonical 3-dimensional QFT to be defined on each Cauchy surface
\Ct, where \Ct
 is defined by
\begin{equation}
  \Ct := \{m \in \M | x^0(m) = t \}
\end{equation}
\M is a globally hyperbolic space-time so these Cauchy surfaces are all
isomorphic. In fact they are all homeomorphic to $ \mathbb{R}^3$ so
${\cal{H}}_t = \F[L^2(\mathbb{R}^3,d^3x)]$ for all times $t$. We define the
history algebra to be (in non-rigorous unsmeared form),
\begin{eqnarray}
    \label{histalg}
  \left[ \phi_{t_1}({\bf{x_1}}) , \phi_{t_2}(\bf{x_2}) \right] &=& 0 \\
  \left[ \pi_{t_1}({\bf{x_1}}) , \pi_{t_2}(\bf{x_2}) \right] &=& 0   \\
  \left[ \phi_{t_1}({\bf{x_1}}) , \pi_{t_2}(\bf{x_2}) \right] &=& i\hbar
                \delta(t_1-t_2) \delta^3(\bf{x_1}-\bf{x_2})
\end{eqnarray}
with ${\bf{x_1}} \in {\cal{C}}_{t_1}$. As shown in \cite{Ish4},
the requirement that the Hamiltonian is self-adjoint is sufficient
to select a representation of this algebra on the history space,
\begin{equation}
 \V^\M := \otimes_{t \in \mathbb{R}}  {\cal{H}}_t  \approx
                                        \F[L^2(\M),d^4x)]
\end{equation}
This representation is defined by the annihilation operator
\begin{equation}
    \label{minkcreat}
  a_t({\bf{x}}) := \frac{1}{\sqrt{2}} \left( K^{1/4}_M \phit + i K^{-1/4}_M
                                        \pit \right)
\end{equation}
where $K_M$ is defined by
$(K_Mf)(t,{\bf{x}}) := (-\nabla_x^2 + m^2)f(t,{\bf{x}})$.
 Equation (\ref{minkcreat}) is a familiar equation in an unusual form. If we
write \phit in
terms of \ak \,  and \adk \, (defined as the 3 dimensional Fourier transforms
of \ax \, and \adx \, respectively) then we have
\begin{equation}
  \phit = \int \frac{d^3k}{(2\omega_k)^{1/2}} ( e^{i{\bf{k.x}}} \adk +
                                e^{-i{\bf{k.x}}} \ak )
\end{equation}
 However, we must not let the familiar form of these equations make us forget
that we are dealing with a history theory. In particular we must
remember that the \phit operator is in the \emph{Schr\"odinger}
picture and the $t$ label that it carries represents the time that
a particular proposition is made, \emph{i.e.} it is a
\emph{logical} time quite separate from \emph{dynamical} time. We
can introduce dynamical time  by using a one parameter unitary
group as usual, but this must involve the introduction of a second
time label:
\begin{eqnarray}
 \phi_t(s,{\bf{x}}) &:=& e^{isH}\phi_t({\bf{x}})e^{-isH}     \\
                    &=&  \int \frac{d^3k}{(2\omega_k)^{1/2}} ( e^{i({\bf{k.x}}
                      -\omega_ks)} \adk + e^{-i({\bf{k.x}}-\omega_ks)} \ak
\end{eqnarray}
 where $H:=\int dt H_t \in {\cal{B}}(\V^\M)$ is the time-averaged Hamiltonian.
Another difference with the canonical theory is that
only projection operators have any meaning. Here we will be interested in
propositions about the number of particles in a particular mode
so we now define these:
\begin{equation}
  N_{\bf{k}} := \int dt \, a^\dagger_t({\bf{k}}) a_t({\bf{k}})
\end{equation}
 This operator has a highly degenerate spectrum as vectors of the form
\begin{equation}
  |n^{\bf{k}}_f \rangle := (n!)^{-1/2} \int dt f(t_1 \dots t_n)
        {a_{t_1}}^{\dagger} ({\bf{k}}) \dots {a_{t_n}}^{\dagger} ({\bf{k}})
                                                                | 0^M \rangle
\end{equation}
 are eigenvectors, with eigenvalue $n \in \mathbb{N}$ for all functions $f$.
 This degeneracy is the result of the fact that we
are considering a free theory, so each $N_{\bf{k}}$ is
separately conserved ($[N_{\bf{k}},H]=0$) and must be an integer. Projectors
$P_{n^{\bf{k}}_f}$
which project onto these vectors represent propositions about the average
number of particles in mode $\bf{k}$ in the interval $t \in supp(f)$.
Symmetry implies that the propositions $P_{n^{\bf{k}}_f}$ form a canonical
consistent set and that the probability of these propositions is decided by the
probability in the initial state:
\begin{equation}
    \label{dM}
   d^\M(m^{\bf{k}}_f , n^{\bf{k'}}_g) = \delta_{mn}\delta^3({\bf{k}}-{\bf{k'}})
                                        \rho^M_{m^{\bf{k}}n^{\bf{k'}}}
\end{equation}
for all $f,g$, where $\rho^M \in {\B}({\cal{H}}_{t_0})$ is defined
by its matrix elements:
\begin{equation}
  \rho^M_{m^{\bf{k}}n^{\bf{k}}} := \langle m^{\bf{k}} | \, \rho^M | n^{\bf{k}}
                                                                        \rangle
\end{equation}
and $| n^{\bf{k}} \rangle := (a^\dagger_{t_0}({\bf{k}}))^n | 0^M
\rangle$.

 We can write the decoherence functional in the form
\begin{equation}
  d^\M(m^{\bf{k}}_f , n^{\bf{k'}}_g) = Tr_{{\cal{V}}^{\M} \otimes
        {\cal{V}}^{\M}}[P_{m^{\bf{k}}_f} \otimes P_{n^{\bf{k'}}_g} \Z^\M ]
\end{equation}
 if $\Z^\M \in {\cal{B}}({\cal{V}}^{\M} \otimes {\cal{V}}^{\M})$ satisfies
\begin{equation}
	\label{zm}
  \langle m^{\bf{k_1}}_f n^{\bf{k_2}}_g | \Z^\M | m'^{\bf{\,k'_1}}_{f'}
n'^{\bf{\,k'_2}}_{g'} \rangle = \delta_{mn}
\delta_{nm'}\delta_{m'n'}  
\delta({\bf{k_1}}-{\bf{k_2}})
\delta({\bf{k_2}}-{\bf{k'_1}}) \delta({\bf{k'_1}}-{\bf{k'_2}})
\rho_{m^{\bf{k_1}}n^{\bf{k_1}}}
\end{equation}
for all $f, f', g, g'$,
 which only defines $\Z^\M$ up to a unitary transformation as before.

\subsection{Canonical QFT on Rindler space-time}

 Consider an observer accelerating with constant acceleration, $\alpha$, 
through \M. Let
$\xi^\mu$ denote the coordinates in which this observer is at rest. Then
$\xi^\mu$ are  related to the coordinates $x^\mu$ by
\begin{equation}
 (x^1)^2-(x^0)^2 = (\xi^1)^2  \; ,  \; x^0/x^1 = tanh(\alpha \xi^0)  \; , \;
        x^2 = \xi^2  \; , \; x^3 = \xi^3
\end{equation}
 So constantly accelerating observers follow hyperbolae in \M.
These hyperbolae split into 2 sets depending on the sign of
$\xi^1$. Rindler space, \R \, , is defined to be the space covered
by the coordinates $\xi^\mu$ with $\xi^1>0$. It corresponds to the
wedge $x>|t|$ in ordinary Minkowski coordinates. Similarly, \cL \,
is defined to be the space covered by $\xi^\mu$ with $\xi^1<0$. It
corresponds to the the wedge $x<|-t|$.
 The metric in these coordinates takes the form,
\begin{equation}
	\label{metric}
 ds^2 := g_{\mu\nu} d\xi^\mu d\xi^\nu = (\alpha \xi^1)^2(d\xi^0)^2 
			- (d\xi^1)^2 -(d\xi^2)^2 - (d\xi^3)^2
\end{equation}

 The vector $\partial_{\xi^0}$ is a globally time-like Killing vector field in
\R. Therefore \R is globally hyperbolic and we can formulate QFT canonically by
using $\partial_{\xi^0}$ to select a particular representation of the canonical
commutation relations. On non-globally hyperbolic space-times there is no 
globally time-like vector field and therefore no way to select one of the 
infinite number of unitarily inequivalent representations. This is the major 
difficulty in the theory of QFT in curved spaces. However, this does not 
concern us here and we proceed by solving the classical Klein-Gordon equation 
in curved space-time:
\begin{equation} 
	\label{KG}
  (g^{\mu\nu} \nabla_\mu \nabla_\nu + m^2)\phi^R(\xi) = 0
\end{equation}
Here, $\nabla_\mu$ is the covariant derivative associated with the metric
(\ref{metric}). As shown in \cite{Full}, equation (\ref{KG}) can be reduced to
a Bessel equation with solutions $u^R_\kappa(\xi)$. Following the 
canonical procedure we now second quantise and expand the quantum field in 
terms of creation and annihilation operators,
\begin{equation}
    \phi^R(\xi) := \int \frac{d^3\kappa}{(2\omega_\kappa)^{1/2}} \left(
       u^R_\kappa(\xi) {b^R(\vec{\kappa})}^\dagger + 
	u^R_\kappa(\xi) b^R(\vec{\kappa}) \right)
\end{equation}
  We can write down a similar equation for the field in \cL and because 
${\cal{C}}^{\cL}_\tau \cup {\cal{C}}^{\R}_\tau$ is a Cauchy surface for \M
we can expand the field on \M as :
\begin{eqnarray*}
  \phi(x) = \int \frac{d^3\kappa}{(2\omega_\kappa)^{1/2}} (
b^R(\vec{\kappa}) \overline{u}^R_\kappa(x) + {b^R(\vec{\kappa})}^\dagger 
					\overline{u}^R_\kappa(x) \\
  +  b^L(\vec{\kappa}) \overline{u}^L_\kappa(x) +  {b^L(\vec{\kappa})}^\dagger 
					\overline{u}^L_\kappa(x) )
\end{eqnarray*}
where 
\begin{eqnarray}
  \overline{u}^R_\kappa(x) &:=&  u^R_\kappa(x) \; if \; x \in \R  \\
		&:=&  0 \; \; \mbox{otherwise}
\end{eqnarray}
and similarly for  $\overline{u}^L_\kappa(x)$.

 Unruh \cite{Unruh1} used the analytic properties of the eigenfunctions 
$u^R_\kappa(x)$ to find the Bogoliubov transformation between the 
above expansion and the usual one:
\begin{equation}
 \phi(x) = \int \frac{d^3k}{(2\omega_k)^{1/2}} \left(a({\bf{k}})
e^{ik.x} + a^\dagger({\bf{k}}) e^{-ik.x} 
\right)
\end{equation}
 Unruh showed that the inertial vacuum can be written
as a thermal density matrix in the Fock basis associated with the accelerating
observer. It is this result that leads to the claim that an accelerating 
observer appears to be immersed in a thermal bath.

\subsection{The Histories Approach}

We now formulate QFT on Rindler space-time using the HPO approach and show how
the the Unruh effect appears within the formalism.
  
Firstly we use the time coordinate of our accelerating observer to 
foliate \R with a one parameter family of spacelike Cauchy surfaces \Ctau
 where
\begin{equation}
 \Ctau := \{r \in \R |\xi^0(r) = \tau \}
\end{equation}
 The single time Hilbert space for the theory is then $\Htau :=
\F[L^2(\Ctau,d\mu)]$  where $d\mu(\xi) = (\alpha \xi^1)^{-1}
d^3\xi$ \cite{Full}. The History space is
\begin{equation}
 \V^\R := \otimes_{\tau \in \mathbb{R}} {\cal{H}}_\tau \approx 
					\F[L^2(\R,d\mu \, d\tau)]
\end{equation}
 By analogy with equations (\ref{histalg}) we define the history algebra to be
\begin{eqnarray}
\left[\phi_{\tau_1}({\vec{\xi_1}}),\phi_{\tau_2}(\vec{\xi_2}) \right] &=& 0 \\
\left[ \pi_{\tau_1}({\vec{\xi_1}}) , \pi_{\tau_2}(\vec{\xi_2}) \right] &=& 0 \\
  \left[ \phi_{\tau_1}({\vec{\xi_1}}) , \pi_{\tau_2}(\vec{\xi_2}) \right] &=&
                i\hbar \delta(\tau_1-\tau_2) \delta^3(\vec{\xi_1}-\vec{\xi_2})
\end{eqnarray}
with $\vec{\xi_1} \in {\cal{C}}_{\tau_1}$.

 The Hamiltonian of the real scalar field in \R is
\begin{equation}
    \label{RHam}
   H^R_\tau = \frac{1}{2} \int d^3 \xi \, \alpha \xi^1 \, (\pixi^2 +
   \nabla_\xi \, \phixi.\nabla_\xi \, \phixi + m^2 \phixi^2)
\end{equation}
 where the vector field $\nabla_\xi$ is defined by $\nabla_\xi :=
 \partial_{\xi^1} + \partial_{\xi^2} + \partial_{\xi^3}$, and the dot
 product is taken using the 3-metric on \Ctau; $g^3 = diag(-1,-1,-1)$ .
 Equation (\ref{RHam}) has the same form for all $\tau$ so the
representation of the history algebra in which $H^R_\tau$ is self-adjoint
is isomorphic on each \Htau.
 The commutation relations of the smeared Hamiltonian with \phixi and \pixi
are
\begin{eqnarray}
 \left[H^R_f,\phixi \right] &=& -i\hbar \, \alpha \xi^1 f(\tau) \pixi       \\
 \left[H^R_f,\pixi \right] &=& i\hbar \, f(\tau) K_R \phixi
\end{eqnarray}
where $K_R$ is defined by $(K_Rf)(\tau,{\vec{\xi}}) :=
(-\nabla_\xi(\alpha\xi^1\nabla_\xi) + \alpha\xi^1 m^2)f(\tau,{\vec{\xi}})$.
 Now we can follow the analysis of Isham \emph{et al} \cite{Ish4} to
show that there is a unitary representation of the exponentiated commutation
relations and that therefore the Hamiltonian exists as a self-adjoint operator
in this representation. We can deduce the associated annihilation
operators to be
\begin{equation}
\label{bR}
   \bxi = \frac{1}{\sqrt{2}} \left( K^{1/4}_R \phixi
 + i\frac{\alpha \xi^1}{K^{1/4}_R}      \pixi \right)
\end{equation}
 This defines a particular complexification of the test function space which is
equivalent to a choice of positive and negative frequencies
consistent with the Killing field $\partial_{\xi^0}$. Using these
creation and annihilation operators we can build the Fock basis
for the history theory. These equations can be written in a more
familiar form by taking the spectral transform of the \bxi \, and
\bdxi, that is by expanding them in terms of the eigenfunctions of
$K_R$, $u^R_\kappa(\vec{\xi})$ \footnote{these are just the functions 
$u^R_\kappa(\xi)$, but with the time dependent part set to 1}:
\begin{equation}
        \phixi := \int \frac{d^3\kappa}{(2\omega_\kappa)^{1/2}} \left(
       u^R_\kappa({\vec{\xi}}) \; \bdk + u^R_\kappa({\vec{\xi}}) \; \bk \right)
\end{equation}
 as before.
 There is obviously a strong similarity between the histories version of this
problem and the canonical version. But from the histories perspective the 
result of Unruh shows nothing because a thermal density matrix
is not a projection operator and so has no meaning when defined on $\V^\M$.
 Only elements of ${\cal{P}}({\cal{V}}^\M)$ and ${\cal{P}}({\cal{V}}^\R)$
 are meaningful in a history theory
as these can be considered as propositions about histories,
\emph{i.e.} as elements of $\cal{U}\cal{P}^\M$ and
$\cal{U}\cal{P}^\R$. We have to change our approach so that we are
talking about projectors onto eigenvectors of the average Rindler
particle number operator:
\begin{equation}
  N_{{\bf{\kappa}}} := \int d\tau \, \bdk \bk
\end{equation}
 These vectors are of the form
\begin{equation}
|n^{\bf{\kappa}}_f \rangle := (n!)^{-1/2} \int d\tau f(\tau_1 \dots \tau_2)
           {b^R_{\tau_1}}^{\dagger} ({\vec{\kappa}}) \dots
                {b^R_{\tau_n}}^{\dagger} ({\vec{\kappa}}) | \, \, 0^R \rangle
\end{equation}
and have a degenerate spectrum in the same way as those for the inertial
observer because we are still considering a free theory.
Projectors onto these vectors represent propositions
about the time-averaged number of particles in each mode, as seen by the
accelerating observer.

The space of propositions about possible histories is not the same
for the accelerating observer as for the inertial observer, but
this is not the only difference. The decoherence functional
associated with a quantum system depends on both the initial
conditions and the Hamiltonian. The accelerating observer has a
different Hamiltonian to the inertial observer and so has a
different decoherence functional.

As before, the fact that $[N_{{\bf{\kappa}}},H^R] = 0$ implies that,
\begin{equation}
   d^\R(m^{\bf{\kappa}}_f , n^{\bf{\kappa'}}_g) = \delta_{mn}\delta^3
({\bf{\kappa}}-{\bf{\kappa'}})\rho^R_{m^{\bf{\kappa}}n^{\bf{\kappa}}}
\end{equation}
 for all $f,g$, in notation which parallels that of (\ref{dM}) but now,
$\rho^R \in {\B}({\cal{H}}^R_{\tau_0})$ and
 $| n^{\bf{\kappa}} \rangle \in {\cal{H}}^R_{\tau_0}$ is defined by:
\begin{equation}
 | n^{\bf{\kappa}} \rangle := ({b^R_{\tau_0}}^\dagger({\bf{\kappa}}))^n
                                                | 0^R \rangle
\end{equation}
 We can write this in the form,
\begin{equation}
  d^\R(m^{\bf{\kappa}}_f , n^{\bf{\kappa}}_g) := Tr_{{\cal{V}}^{\R} \otimes
 {\cal{V}}^{\R}}[P_{m^{\bf{\kappa}}_f} \otimes P_{n^{\bf{\kappa}}_g} \Z^\R ]
\end{equation}
 for $\Z^\R \in {\cal{B}}({\cal{V}}^{\R} \otimes {\cal{V}}^{\R})$ defined
similarly to the Minkowski case, (\ref{zm}).
 
\subsection{The Unruh Effect}
 Finally we can see how the Unruh effect arises in the HPO formalism.
Let us consider the situation in the inertial vacuum, \emph{i.e.}
the initial density matrix is
\begin{equation}
 \rho^M_{n^{\bf{k}}n^{\bf{k}}} = \delta_{0n}
\end{equation}
for all $\bf{k} \in \mathbb{R}^3$, where the matrix elements are
taken in the Fock representation associated with the inertial
observer. Note that this density matrix means that the probability
of the inertial observer detecting $n$ particles in any mode is
zero unless $n=0$:
\begin{equation}
  d^\M(m^{\bf{k}}_f , n^{\bf{k'}}_g) = \delta_{mn}\delta({\bf{k}}-{\bf{k'}})
                                                                \delta_{0n}
\end{equation}
The density matrix $\rho^M$ is
defined on some initial Hilbert space ${\cal{H}}_{t_0}$, but we can choose our
Cauchy surfaces so that:
\begin{equation}
  {\cal{H}}_{t_0} = {\cal{H}}^{\cL}_{\tau_0} \otimes {\cal{H}}^{\R}_{\tau_0}
\end{equation}
 Using Unruh's result on this initial Hilbert space we can write the
inertial vacuum as a thermal density matrix in the representation associated
with the accelerating observer. Tracing over ${\cal{H}}^{\cL}_{\tau_0}$ we
obtain the initial condition for the accelerating observer \cite{Unruh}
\begin{equation}
 \rho^R_{n^{\bf{\kappa}}n^{\bf{\kappa}}} = N_\frac{2\pi}{\alpha}
                                                (n \omega_\kappa)
\end{equation}
 where $N_\beta(E)$ is the thermal distribution giving the probability of a
scalar particle having energy $E$ in a heat bath of inverse temperature
$\beta$. Finally,
\begin{equation}
   d^\R(n^{\bf{\kappa}}_f , m^{\bf{\kappa'}}_g) = \delta_{mn}
\delta({\bf{\kappa}}-{\bf{\kappa'})}N_\frac{2\pi}{\alpha}
                                                        (n\omega_\kappa),
\end{equation}
which shows that the accelerating observer detects a thermal spectrum at
inverse temperature $\beta = \frac{2\pi}{\alpha}$, in agreement with the result
of Unruh.

\section{Conclusion}
 We have shown that it is possible to consider average number propositions
within the continuous time HPO formalism. We have postulated a
condition on the decoherence functional which ensures that energy
propositions form a consistent set, as they do in the conventional
theory, and which gives the correct probabilities for such
propositions. This condition is defined for the SHO and for QFT
but can easily be generalised to any system with symmetries as its
construction involves only the matrix
 elements of the initial density matrix in the basis associated with the
symmetry.

We have shown that the HPO scheme allows the construction of QFT
in curved space-time and have re-derived the well-known result of
Unruh within this scheme. In fact, the general nature of the HPO
formalism - in particular its ability to cope with very general
temporal support strucures and the associated non-unitary
evolution - means that it can potentially be used to formulate QFT
on much more general space-times such as non-globally hyperbolic
space-times or those with topology change. This remains a task for
future research.

 Another potentially interesting avenue of research is to attempt to apply the
 formalism to
other problems in conventional QFT such as scattering. Scattering type
questions typically involve propositions such as "there are $n_1$ particles of
type $1$ at time $t_1$ and then $n_2$ particles of type $2$ at time $t_2$". We
cannot pose such questions in the formalism as presented here because we cannot
embed discrete time propositions into the continuous time history space. The
best we can do is
 to use propositions with support in a neighbourhood of $t_1$ and $t_2$ which
we can choose arbitrarily small. Non-trivial scattering questions
necessarily involve interactions and we haven't considered these
here, but in principle there is no reason why perturbation theory
could not be developed.

\section{Acknowledgements}
 I would like to thank Chris Isham for many useful discussions.

\end{document}